\documentclass[prd,aps,twocolumn,preprintnumbers, showpacs, nofootinbib,superscriptaddress,notitlepage]{revtex4-1}
\usepackage{amsmath,graphicx,epsfig,amssymb}
\usepackage{inputenc}
\usepackage[usenames]{color}
\usepackage{ulem} 
\usepackage{bigstrut}
\usepackage{slashed}
\usepackage{multirow}
\usepackage{subfigure}
\usepackage{dsfont}
\usepackage{xcolor}
\usepackage{hyperref}
\usepackage{graphicx}
\usepackage{subcaption} 

\allowdisplaybreaks

\begin{document}

\title{Solving the Inverse Source Problem in Femtoscopy with a Toy Model}

\author{Ao-Sheng Xiong}
\affiliation{Frontiers Science Center for Rare Isotopes, and School of Nuclear Science and Technology, Lanzhou University, Lanzhou 730000, China}

\author{Qi-Wei Yuan}
\affiliation{Frontiers Science Center for Rare Isotopes, and School of Nuclear Science and Technology, Lanzhou University, Lanzhou 730000, China}

\author{Ming-Zhu Liu}
\email[Corresponding author: ]{liumz@lzu.edu.cn}
\affiliation{Frontiers Science Center for Rare Isotopes, and School of Nuclear Science and Technology, Lanzhou University, Lanzhou 730000, China}

\author{Fu-Sheng Yu}
\email[Corresponding author: ]{yufsh@lzu.edu.cn}
\affiliation{Frontiers Science Center for Rare Isotopes, and School of Nuclear Science and Technology, Lanzhou University, Lanzhou 730000, China}

\author{Zhi-Wei Liu}
\email[Corresponding author: ]{liuzhw@szu.edu.cn}
\affiliation{Institute for Advanced Study in Nuclear Energy \& Safety, College of Physics and Optoelectronic Engineering, Shenzhen University, Shenzhen 518060, China}
\affiliation{Shenzhen Key Laboratory of Nuclear and Radiation Safety, Shenzhen 518060, China}

\author{Li-Sheng Geng}
\email[Corresponding author: ]{lisheng.geng@buaa.edu.cn}
\affiliation{School of Physics, Beihang University, Beijing 102206, China}
\affiliation{
Sino-French Carbon Neutrality Research Center, \'Ecole Centrale de P\'ekin/School
of General Engineering, Beihang University, 
Beijing 100191, China}
\affiliation{Peng Huanwu Collaborative Center for Research and Education, Beihang University, Beijing 100191, China}
\affiliation{Beijing Key Laboratory of Advanced Nuclear Materials and Physics, Beihang University, Beijing 100191, China }
\affiliation{Southern Center for Nuclear-Science Theory (SCNT), Institute of Modern Physics, Chinese Academy of Sciences, Huizhou 516000, China}

\begin{abstract}

Hadron-hadron interactions, as a non-perturbative effect, play a significant role in understanding phenomenological problems in particle physics.   
Femtoscopy is a powerful tool in heavy-ion collision experiments, enabling the extraction of hadron-hadron interactions via momentum-correlation functions (CFs). These CFs are generally factorized into a convolution of source functions and hadron-hadron wave functions, with the latter encoding information about hadron-hadron interactions. However,  source functions remain ambiguous and are commonly approximated by a Gaussian form.
Reconstructing source functions from experimental correlation data constitutes an ``inverse problem." To address it, we propose a toy model based on the Tikhonov regularization. Employing a square potential well of four distinct potential strengths, we calculate the CFs for inputs of a Gaussian source function and its mixed form. The obtained CFs are subsequently used to reconstruct the source functions via the Tikhonov regularization.   
Our results demonstrate that the Gaussian source function can be successfully reconstructed, indicating the potential of this approach for extracting realistic source functions of hadron pairs of interest in the future. 

\end{abstract}
\maketitle
\section{Introduction}

As a residual force of QCD, hadron-hadron interactions play a significant role across various domains of particle physics, ranging from hadron structure and strong decays to weak decays of heavy-flavored hadrons.
Recent studies on exotic states further highlight the importance of hadron-hadron interactions in understanding their properties, suggesting substantial molecular-hadronic components.
Theoretical predictions for the ratios of strong decays  $\mathcal{B}[\psi(2S) \to \rho \pi] / \mathcal{B}[\psi(1S) \to \rho \pi]$ exhibit notable discrepancies with experimental data, a long-standing issue known as the ``$\rho$–$\pi$ puzzle"~\cite{Franklin:1983ve}.
A conventional explanation attributes this puzzle to final-state interactions (FSIs), i.e., hadron-hadron interactions, considered essential for resolving the observed discrepancy~\cite{Li:1996yn,Zhao:2010mm,Zhao:2010zzv,Wang:2012mf}. FSIs also play a crucial role in the weak decays of heavy hadrons, significantly enhancing the branching fractions of nonfactorizable decays~\cite{Colangelo:2002mj,Cheng:2004ru,Oset:2016lyh,Miyahara:2015cja,Ling:2021qzl,Hsiao:2019ait,Jia:2024pyb} and giving rise to substantial CP violation signals in these processes~\cite{Duan:2024zjv,Wang:2024oyi}. Therefore, hadron-hadron interactions are an essential input for phenomenological studies across particle physics, crucial for addressing problems associated with non-perturbative effects.

Currently, advances in lattice QCD have enabled detailed investigations of hadron–hadron interactions directly from first principles of QCD. In recent years, lattice QCD has successfully reproduced the nucleon–nucleon potential~\cite{Ishii:2006ec}, demonstrating its capability to explore interactions between hadrons.
Furthermore, invariant mass distributions are effective experimental observables for extracting information about hadron–hadron interactions.  Cabibbo proposes a notable example to determine the $\pi\pi$ scattering length from the $\pi^0\pi^0$ invariant mass distribution in the decay $K^+ \to \pi^+\pi^0\pi^0$~\cite{Cabibbo:2004gq}. This is a method for determining dominant hadron-hadron interactions. Meanwhile, discoveries of exotic states in invariant-mass distributions shed light on interactions between nearby hadron pairs.  
Recently, femtoscopy, a technique that analyzes momentum correlations between particles produced in high-energy collisions, has emerged as a powerful alternative method for probing the strong interaction~\cite{Fabbietti:2020bfg}. By measuring momentum-correlation functions (CFs), femtoscopy provides valuable insights into hadronic interactions. A particular strength of this approach lies in its ability to investigate hadron-hadron interactions involving unstable hadrons, which are not accessible via conventional scattering experiments~\cite{STAR:2014dcy,STAR:2015kha,ALICE:2019gcn,ALICE:2019hdt,ALICE:2020mfd,ALICE:2021cpv}. This unique capability has stimulated significant theoretical developments~\cite{Morita:2014kza,Morita:2016auo,Ohnishi:2016elb,Haidenbauer:2018jvl,Morita:2019rph,Kamiya:2019uiw,Ogata:2021mbo,Kamiya:2021hdb,Haidenbauer:2021zvr,Liu:2024nac}.

The momentum CFs are described by the Koonin-Pratt (KP) formula~\cite{Koonin:1977fh,Pratt:1990zq}, which incorporates two essential components:  1) a particle-emitting source, characterizing the spatial distribution of hadron emission in relativistic heavy-ion collisions; 2) the scattering wave function, encoding the final-state interactions between hadron pairs. In general, the hadron-hadron interaction can be extracted from  CFs using a known source function. Therefore, a precise source function is crucial for precisely extracting these interactions.  The discussions on the particle-emitting source have been intensive~\cite{ALICE:2020ibs,ALICE:2023sjd,Mihaylov:2023pyl,Xu:2024dnd,Wang:2024bpl}. Sources are often modeled as Gaussian functions. In this case, the characteristic source size produced in proton-proton collisions at LHC energies is about $1$ fm~\cite{ALICE:2020mfd}, while $3-5$~fm  in nucleus-nucleus central collisions~\cite{STAR:2014dcy}. It is worth noting that an effective Gaussian source has been successfully applied in various experimental analyses~\cite{ALICE:2019hdt,ALICE:2020mfd}. Alternatively, a Cauchy source has also been employed to characterize source functions.  
In a recent study, Wang et al. employed machine learning to extract the source function of the proton–proton system, using precise wave functions and experimental data~\cite{Wang:2024bpl}.
 However, the so-determined source function deviates from a Gaussian form. Indeed, the task of reconstructing source functions from experimental correlation data falls within the category of an ``inverse problem", offering a novel perspective on extracting source functions.  
 
 The theory of inverse problems, a cornerstone of applied mathematics, provides a rigorous framework for reconstructing unknown quantities from indirect measurements and has been extensively applied in fields such as geophysical exploration and medical imaging~\cite{Application_in_sci_and_eng}.  The reconstruction of source distributions from measured correlation functions was pioneered by Brown and Danielewicz through the so-called image technique~\cite{Brown:1997ku,Brown:1997sn}. This method employs optimized discretization to address the inherent instability of the inverse problem. The fundamental principle of the optimized discretization approach is to adjust the sizes of the radial bins $r_{j}$ in order to minimize the relative error in the reconstructed source. Then motivated by the problem in optical imaging,  Danielewicz et al., treat the imaging problem as an Optical Deblurring Problem solved via Bayesian Iteration~\cite{Nzabahimana:2023tab,Tam:2025mkk}. This iterative Bayesian method leverages the natural positivity of both the source and the kernel, and successfully incorporates detector resolution effects through convolution.  While both methods have proven powerful, they rely on either optimized discretization strategies or iterative Bayesian frameworks.  In this work, we explore an alternative path: we formulate the source reconstruction as a classical inverse problem and employ Tikhonov regularization, a mathematically rigorous technique that stabilizes the inversion through a penalty term\footnote{It is important to distinguish our work from studies that also invoke the concept of an inverse problem but employ it as a parameter-fitting strategy. For instance, Refs.~\cite{Ikeno:2023ojl,Albaladejo:2023wmv,Molina:2023jov,Li:2024tvo} constrain parametrized wave functions by fitting correlation data to extract scattering observables.

 In our approach, the problem of reconstructing the source function via the KP formula is naturally cast as a Fredholm integral equation of the first kind, a classical formulation for which robust numerical methods have been developed~\cite{Kirsch_2011,Engl_1996,Tik_for_Fredholm_equation}. To solve it, we employ Tikhonov regularization, a widely used technique valued for its mathematical rigor, independence from tunable parameters or phenomenological assumptions, and proven effectiveness across a broad range of applications~\cite{Regularization_tools,Modern_regularization_methods}.  This approach has proven valuable in specialized physical applications, including the extraction of hadronic spectral functions and the determination of decay constants and distribution amplitudes~\cite{Li:2020xrz,Xiong:2022uwj,Xiong:2022uwj,Li:2020xrz,Li:2020ejs,Li:2021cnd,Zhao:2024drr,Xiong:2025obq,Ling:2025olz,Mutuk:2024jvv,Mutuk:2025lak}.  

}

 This paper is structured as follows. Following the methodology of Ref.~\cite{Liu:2023uly}, we first generate momentum correlation functions from wave functions derived from a square-well potential with four distinct interaction strengths—repulsive, weakly attractive, moderately attractive, and strongly attractive. These are combined with Gaussian-type source functions and their mixtures. Subsequently, we attempt to reconstruct these source functions by solving the inverse problem, using the resampled correlation functions (with $1\%$ and $10\%$ uncertainties) along with the corresponding wave functions. Section II introduces the formalism for computing correlation functions using both a square-well potential and a Gaussian-type source function, as well as the approach to the inverse problem. Our numerical results and a detailed discussion are presented in Section III. Finally, we provide a summary and outlook in the last section.
\section{ THEORETICAL FRAMEWORK }

Since the current parameterization of source functions is based on specific assumptions, it is essential to develop a new approach to obtain them in a model-independent manner. In this work, we propose a rigorous mathematical framework for computing source functions. Given the momentum CFs and scattering wave functions, the source function can be reconstructed via an inverse problem. This section provides a concise introduction to the theoretical framework of momentum CFs and the inverse problem method.

\subsection{Source Functions and Wave Functions}
The two hadron momentum CFs are computed by the KP formula~\cite{Koonin:1977fh,Pratt:1990zq,Bauer:1992ffu}  
\begin{equation}
    \begin{aligned}
        C(k)=\int d^3r S_{12}(r) \left| \Psi^{(-)} (r, k) \right|^2\,,
    \end{aligned} \label{KP_formula}
\end{equation}
where $S_{12}(r)$ denotes the source function, and $\Psi^{(-)} (r, k)$ denotes the  relative wave
function with the relative coordinate $r$ and the relative momentum $k=(m_2 p_1-m_1 p_2)/(m_1+m_2)$ in the center-of-mass frame. The source function characterizes the spatial distribution of the emission source, and the wave function describes the final-state hadron-hadron interactions. 

In the following, we adopt a Gaussian-type function to represent the source function,
\begin{equation}
    \begin{aligned}
        S_{12}(r) = \frac{1}{(2\sqrt{\pi}R)^3} \exp{(-r^2/4R^2)}\,,
    \end{aligned}
\end{equation}
where $R$ is the effective radius of the source.

The  outgoing  wave function  of an $S$-wave interaction    is generally  expressed as~\cite{Ohnishi:2016elb}
\begin{equation}
    \begin{aligned}
        \Psi_S^{(-)}(r,k) = e^{ikr}-j_0(kr)+\psi_0(r,k), 
    \end{aligned}
\end{equation}
where the Bessel function $j_0$ represents the component of the orbital angular momentum $l=0$ in the free wave function, while $\psi_0$ represents the $S$-wave scattering wave function after the final-state interaction correction.  The scattering wave function follows the asymptotic behavior, 
\begin{equation}
    \begin{aligned}
        \psi_0(r,k)  \to \frac{1}{2ikr}(e^{ikr}-e^{-2i \delta}e^{-ikr}){\quad(r \to \infty)}, 
    \end{aligned}
\end{equation}
where  $\delta$ represents the phase shift. Finally, the momentum CFs are written as 
\begin{equation}
    \begin{aligned}
        C(k) = 1 + \int_0^\infty 4\pi r^2 dr S_{12}(r) (\left| \psi_0(r,k)   \right|^2 - \left| j_0(kr) \right|^2). 
    \end{aligned}
\end{equation}
In general, the scattering wave function can be obtained by solving the Schr\"odinger equation in coordinate space or the Lippmann-Schwinger equation in momentum space~\cite{Morita:2014kza,Mihaylov:2018rva,Haidenbauer:2018jvl}. In the following, we employ the Schr\"odinger equation to compute the scattering wave functions.  As indicated in Ref.~\cite{Epelbaum:2025aan}, the off-shell ambiguity in strong interactions is generally associated with the short‑distance behavior of relative wave functions, which introduces theoretical uncertainties into calculations of CFs. However, the off‑shell ambiguities are not particularly significant in practical physics. Some quantitative studies have indeed shown that such off‑shell effects remain relatively mild within realistic interaction models for the two‑body problem~\cite{Gobel:2025afq,Molina:2025lzw}.  Hence, the off‑shell ambiguity certainly exists; its impact on two‑body systems may not be very large. In our toy model, we assume the wave function is already known, which does not involve the issue raised by Ref.~\cite{Epelbaum:2025aan}.

\subsection{Inverse problem approach}

 Before detailing the methodology of inverse problems, it is useful to recall the notion of a forward problem. A forward problem consists of predicting results from known causes with a given model. In contrast, an inverse problem aims to infer the causes from observed results and the model. As an illustration, the KP formula corresponds to a forward problem: given the wave function and the source function, the correlation function is obtained by direct integration, which is generally straightforward. Conversely, the inverse problem that reconstructing the source function from the measured correlation function and the wave function is often challenging because it is typically ill-posed.

A central aspect of treating inverse problems is the study of their ill-posedness. Mathematically, a problem is termed well-posed if it satisfies three conditions simultaneously:  existence, uniqueness, and stability of the solution ~\cite{Kirsch_2011,Engl_1996}. Existence requires at least one solution; uniqueness ensures at most one solution; and stability means that the solution depends continuously on the input data. If any of these three conditions is violated, the problem is ill-posed.

For a deterministic system in which the wave function is known precisely (e.g., the proton-proton system with accurately prescribed interactions), the inverse problem of extracting the source function from an experimentally measured correlation function is ill-posed. Specifically, it fails to satisfy the stability condition: an arbitrarily small perturbation in the input can produce arbitrarily large changes in the reconstructed source. To trace the origin of this instability, we discretize the KP equation using a rectangular quadrature rule, which leads to a linear algebraic system
\begin{equation}
    \begin{aligned}
        K_{m \times n}S_n=C_m
    \end{aligned}
\end{equation}
where the matrix $K_{m\times n}$ is built from the values of $\left| \Psi^{(-)} (r_i, k_j) \right|^2$, while $S_n$ and $C_m$ denote the discrete source and correlation functions, respectively. Here, $n$ and $m$ are the dimensions of the source and correlation vectors. Performing a singular value decomposition (SVD) yields
\begin{equation}
    \begin{aligned}
        K = U \Sigma V^T = \sum_{i=1}^{n} \sigma_i u_i  v^T_i\,,
    \end{aligned}
\end{equation}
with $U = [u_1, \ldots, u_n] \in \mathbb{R}^{m \times n}$ and $V=[v_1, \ldots, v_n] \in \mathbb{R}^{n \times n}$ containing orthonormal singular vectors ($U^T U = V^T V = I_n$). The diagonal matrix $\Sigma=\text{diag}(\sigma_1, \ldots, \sigma_n)$ holds the singular values ordered as $\sigma_1 \geq \cdots \geq \sigma_n \geq 0$~\cite{Singular_values_and_condition_numbers}. The formal solution of $C=KS$ is then
\begin{equation}
    \begin{aligned}
        S = \sum_{i=1}^n \frac{u_i^T C}{\sigma_i} v_i.
    \end{aligned}\label{SVD solution}
\end{equation}
which, notably, is unique irrespective of the relative sizes of $m$ and $n$~\cite{Singular_values_and_condition_numbers}. The numerical stability of the inversion is governed by the condition number $\kappa(K)=\sigma_1/\sigma_n$. When $\sigma_1$ and $\sigma_n$ differ by many orders of magnitude, the large condition number induces severe numerical instability: high-frequency noise in the data is dramatically amplified in the reconstructed solution. Such ill‑conditioning is inherent to the discretization of integral equations with smooth kernels: the resulting matrix $K$ has rows (or columns) that are nearly linearly dependent, and this near‑linear dependence forces the singular values to decay rapidly. Besides, using SVD analysis, we quantify the instability
\begin{equation}
    \begin{aligned}
        \left\| S^\epsilon-S_t \right\|_{l^2}^2 = \sum_{i=1}^n \left( \frac{u_i^T (C^\epsilon-C)}{\sigma_i} v_i \right)^2 \to \infty \,,
    \end{aligned}
\end{equation}
where $C$ denotes the exact data, $S_t$ the corresponding true source function, and $C^\epsilon=C+\epsilon$ the measured correlation function with an error level $\epsilon$. Here $\| \cdot \|_{l^2}$ is the standard $l^2$ norm on the sequence space and $S^\epsilon$ is the solution obtained from the noisy data. As the discretization is refined, the inverses of the singular values $\sigma_i$ become unbounded. 

  The existence of a solution is often taken for granted based on the underlying physical picture. A rigorous uniqueness analysis of the inverse problem associated with the KP equation would require a functional-analytic framework, relying on an analytic kernel and continuous, noise-free data—conditions that are rarely met in practice. In particular, the matrix $K$ may have a nontrivial null space, implying that certain components of the source function cannot be recovered without additional assumptions. Therefore, strict uniqueness cannot be guaranteed in the present setting. Nevertheless, it is common to assume that the corresponding continuous problem admits a unique solution~\cite{Gobel:2025afq,Molina:2025lzw}. In this work, our goal is not to establish uniqueness rigorously, but to demonstrate the practical feasibility of reconstructing the source function from finite and noisy data~\cite{IP_Discrete, IP_Discrete_II}. As shown in Sec.~III, the reconstructed solutions capture the essential features of the exact source, indicating that the method is effective in practice even without guaranteed uniqueness. Thus, uniqueness is not the central limitation; rather, the main challenge is numerical instability, which persists even with large data sets or a continuously treated correlation function.


To address this ill-posedness, regularization methods are employed. In mathematics, regularization refers to a family of techniques that transform an originally ill-posed problem into an approximately well-posed one, thereby ensuring the existence, uniqueness, and stability of an approximate solution~\cite{Kirsch_2011,Engl_1996}. While the use of “regularization” in physics—such as in the removal of divergences in loop integrals—differs in context from its mathematical counterpart for ill-posed inverse problems, both share the same essential objective: to achieve a well-defined and numerically stable formulation.

In this work, we employ the Tikhonov regularization scheme, a well-established approach that provides a rigorous framework for solving ill-posed inverse problems. The regularized solution $S_\alpha^\epsilon$ is obtained by minimizing the Tikhonov functional
\begin{equation}
    \begin{aligned}
        S^\epsilon_\alpha = \mathop {\arg \min }\limits_{S \in l^2} \left \| K S-C^\epsilon \right\|_{l^2}^2+\alpha \left \| L S \right\|_{l^2}^2\,,
    \end{aligned} \label{Tikhonov_functional}
\end{equation}
where $\alpha>0$ is the regularization parameter. This method, pioneered by Tikhonov, effectively stabilizes the inversion by penalizing large oscillations in the solution, and it underlies many contemporary regularization techniques used in fields such as~\cite{Modern_regularization_methods}. The first term enforces fidelity to the data $C^\epsilon$, while the second term penalizes oscillations to stabilize the solution~\cite{Xiong:2022uwj,Xiong:2025obq,Ling:2025olz}. The matrix $L$ incorporates prior constraints on the solution such as smoothness, positivity, or normalization constraints; here, for simplicity, we choose it to be a first‑derivative operator to enforce smoothness, which is sufficient for the present analysis. Its discrete $(n-1)\times n$ form is
\begin{equation}
L = \begin{pmatrix}
1 & -1 & 0 & \cdots & 0 \\
0 & 1 & -1 & \ddots & \vdots \\
\vdots & \ddots & \ddots & \ddots & 0 \\
0 & \cdots & 0 & 1 & -1
\end{pmatrix}\,.
\end{equation}

Performing the variation of the Tikhonov functional leads to
\begin{equation}
    \begin{aligned}
        (K^\mathsf{T} K + \alpha L^\mathsf{T} L) S_\alpha^\epsilon = K^\mathsf{T} C^\epsilon
    \end{aligned} \label{Tikhonov variation}
\end{equation}
where $\mathsf{T}$ denotes the matrix transpose. Eq.\,(\ref{Tikhonov variation}) illustrates how Tikhonov regularization stabilizes the inversion: the addition of $\alpha L^\mathsf{T}L$ effectively reduces the condition number of the system. In particular, the regularized system maps $\sigma_i$ to $\sigma_i + \alpha$ (for an appropriate $L$), so that the original condition number $\sigma_1/\sigma_n$ becomes
\begin{equation}
    \begin{aligned}
        \frac{\sigma_1}{\sigma_n} \to \frac{\sigma_1+\alpha}{\sigma_n+\alpha} \approx \frac{\sigma_1+\alpha}{\alpha}
    \end{aligned}
\end{equation}
This clearly suppresses the amplification of noise in the small‑singular‑value directions and renders the problem well‑posed.

A crucial aspect of the Tikhonov regularization is the choice of the regularization parameter $\alpha$.  The Tikhonov functional illustrates that the choice of the regularization parameter $\alpha$ governs a trade-off between the fidelity term and the regularization term. If $\alpha$ is excessively large, the regularization term dominates, thereby forcing the solution toward zero. Conversely, if $\alpha$ is too small, the fidelity term prevails and the inherent ill-posedness of the inverse problem re-emerges. Empirically, $\alpha$ often exhibits a broad plateau region that aids its robust selection.  Numerous methods have been developed in inverse problems for determining the regularization parameter, including the L-curve criterion, the quasi-optimality criterion and the discrepancy principle, among others~\cite{Kirsch_2011,Engl_1996}.  In this work, we employ the well-used L-curve criterion~\cite{The_posterier_parameter_choice, HPC_1993}, which leverages the systematic trade-off between the residual $\|K S_\alpha^\epsilon - C^\epsilon\|_{l^2}$ and the penalty $\| L S_\alpha^\epsilon \|_{l^2}$: as $\alpha$ decreases, the residual decreases while the penalty increases. The L-curve is a log–log plot of these two quantities, and the optimal $\alpha$ is identified at its characteristic corner. This approach is fully deterministic and ensures that the inversion procedure involves no arbitrarily tuned parameters.

The reconstruction of the source function via the Tikhonov regularization is governed by a rigorous mathematical framework that establishes two fundamental properties: numerical stability (well-posedness) and theoretical convergence. Specifically, as the noise level in the data tends to zero, 
the regularized solution $S_\alpha^\epsilon$ is guaranteed to converge to the true solution $S_t$. This result can be stated as follows:
\begin{equation}
    \begin{aligned}
        \|S^\epsilon_\alpha-S_t\|_{l^2} \to 0, \, \epsilon \to 0\,.
    \end{aligned} \label{Tik_convergence}
\end{equation}
A comprehensive theoretical treatment, including proofs of well‑posedness and convergence, is provided in Refs.~\cite{Kirsch_2011,Engl_1996,Tik_for_Fredholm_equation,Regularization_tools,Modern_regularization_methods}. 

\section{Numerical results}
In this section, we assess the performance of the Tikhonov regularization for the inverse source problem in Femtoscopy using a toy model. We consider four distinct potential strengths and compute the corresponding wave functions by solving the Schr\"odinger equation. The true source functions, $S_t(r)$, are modeled as both single and mixed Gaussian forms. Using these wave functions and different source functions, we calculate the CFs $C_t(k)$ via the KP formula. The perturbed CFs $C^\epsilon(k)$ are generated by the introduction of $10\%$ and $1\%$ random errors into the $C_t(k)$, simulating realistic experimental uncertainties.   This noisy $C^\epsilon(k)$ then serves as the input to the Tikhonov regularization, which yields the reconstructed source function $S^\epsilon_\alpha(r)$. A comparison of $S^\epsilon_\alpha(r)$ with the benchmark $S_t(r)$ allows us to evaluate the effectiveness of the Tikhonov regularization.

We now outline the potentials and source functions used in this work. We employ a square-well potential, chosen for its analytic wave function and ability to capture generic features of the interaction through parameter adjustments. The potential is defined as:
\begin{equation}
    \begin{aligned}
        V(r)=V_0 \theta(d-r)\,,
    \end{aligned}
\end{equation}
where $V_0$ and $d$ represent the strength and range of the potential, respectively. Following Ref.~\cite{Liu:2023uly}, we fix the range parameter at $d=2.5 \, \mathrm{fm}$, and take $V_0=25,-10,-25,-75 \, \mathrm{MeV}$, corresponding to repulsive, weakly attractive, moderately attractive, and strongly attractive potentials, respectively. 

The source functions fall into four types: the first two are single Gaussians, and the remaining two are Gaussian mixtures. This selection allows us to study both standard and more general source shapes. Explicitly, the sources are given by $S_1(r,R)$, $S_2(r,R)$ (single Gaussians), and $S_3(r,R)$, $S_4(r,R)$ (mixed Gaussians of the form $wS(r,R_1)+(1-w)S(r,R_2)$, where $w$ is a weighting factor). The values of relevant parameters are summarized in Table \,\ref{tab:source_params}. 

\begin{table}[htbp]
    \centering
    \caption{Parameter configurations for the source functions.}
    \label{tab:source_params}
    \begin{tabular}{ccc}
        \hline \hline
        ~~~~Source~~~~ & ~~~Type~~~ & ~~~Parameters~~~ \\
        \hline
        $S_1(r,R)$ & Single & $R = 1$~ fm \\
        $S_2(r,R)$ & Single & $R = 1.5$~ fm \\
        $S_3(r,R)$ & Mixed  & $w = 0.2$, $R_1 = 1$~fm, $R_2 = 3$~fm \\
        $S_4(r,R)$ & Mixed  & $w = 0.1$, $R_1 = 1$~fm, $R_2 = 3$~fm \\
        \hline \hline
    \end{tabular}
\end{table}

\begin{figure*}[htbp]
\centering
\includegraphics[width=0.65\linewidth]{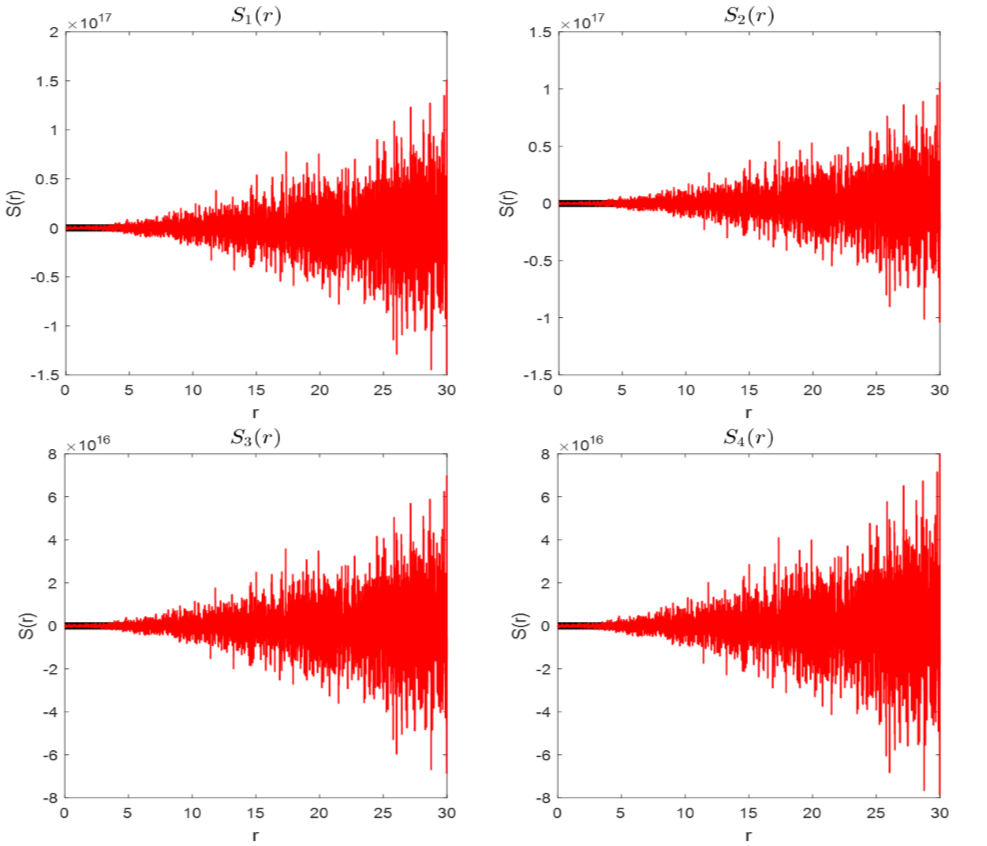}
\caption{Unregularized solutions (red curves) for the four sources, exhibiting unstable reconstructions that deviate from the benchmark solutions (black curves) by 17 to 18 orders of magnitude.}
\label{fig:No_regularization}
\end{figure*}

To numerically illustrate the inherent ill-posedness of reconstructing the source function from CFs, we begin by applying the standard SVD method  given by Eq.\,(\ref{SVD solution}) without regularization. As a representative case,  we consider the potential  $V_0=-10\,\mathrm{MeV}$ with a $1\%$ uncertainty.  Preforming an SVD on the corresponding kernel matrix $K$ yields singular values
\begin{equation}
    \begin{aligned}
        \sigma_1=35.1727 \geq \cdots \geq \sigma_n = 2.6605 \times 10^{-15}
    \end{aligned}
\end{equation}
which span more than 15 orders of magnitude. The resulting condition number of $K$ is therefore extremely large, confirming that the underlying inverse problem is severely ill-conditioned. As shown in Fig.\,\ref{fig:No_regularization}, the unregularized reconstructions (red curve) exhibit pronounced unphysical oscillations and deviate significantly from the benchmark (black curve). This oscillatory behavior provides direct numerical evidence of the ill-posedness of the inverse problem, indicating  that even minor input perturbations can induce substantial instabilities in the solution.  Critically, this instability is fundamental: it arises irrespective of whether the number of data points is larger, smaller, or equal to the number of source degrees of freedom, and it persists even for continuous input data~\cite{Kirsch_2011,Engl_1996}. These results underscore the need to employ the Tikhonov regularization to mitigate numerical artifacts. 

\begin{figure*}[htbp]
\centering
\includegraphics[width=0.65\linewidth]{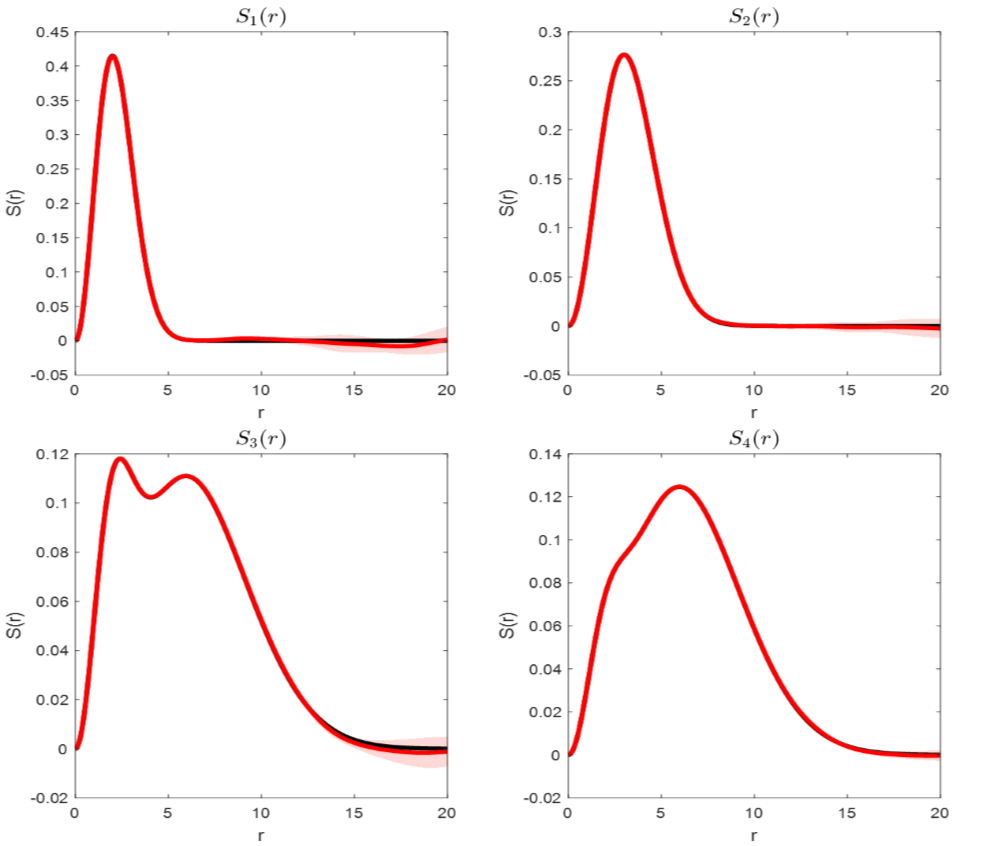}
\caption{Reconstructed solutions $S_\alpha^\epsilon$ (red lines) and the associated error bands (pink shaded areas) for $V_0=-10\,\mathrm{MeV}$ with $1\%$ uncertainty, compared to the benchmarks $S_t$ (black line).  }
\label{fig:wave_m10_with_01}
\end{figure*}

\begin{figure*}[htbp]
\centering
\includegraphics[width=0.65\linewidth]{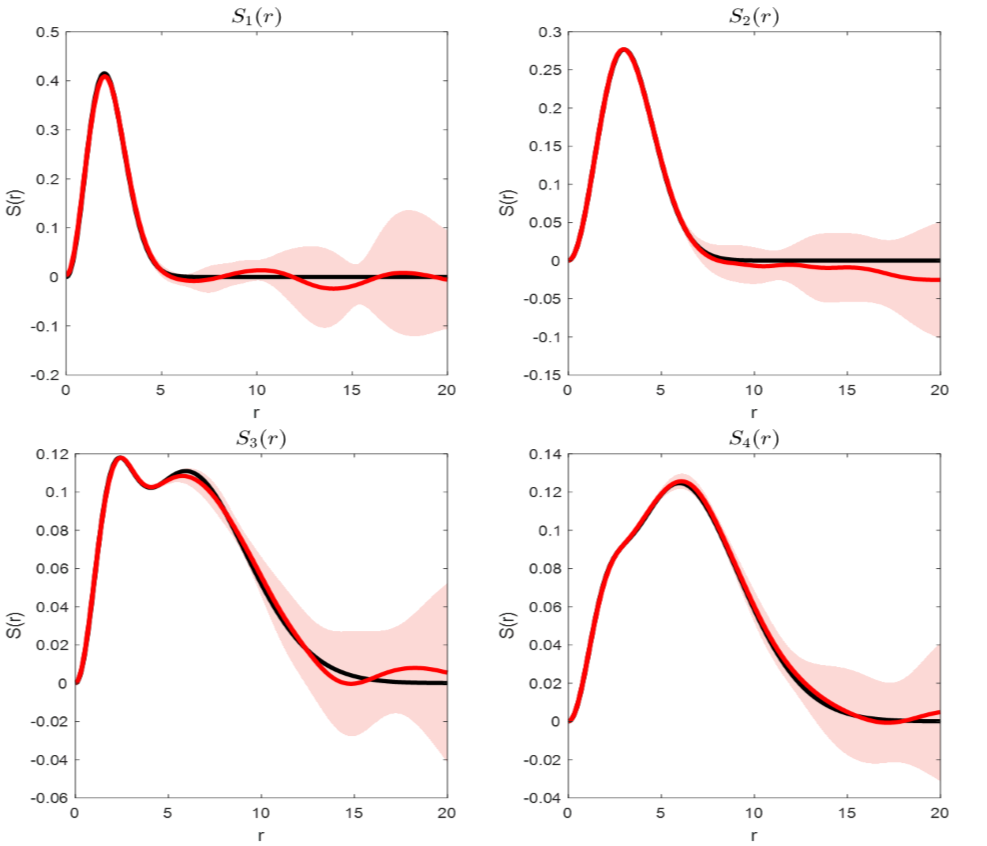}
\caption{Reconstructed solutions $S_\alpha^\epsilon$ (red lines) and the associated error bands (pink shaded areas) for $V_0=-10\,\mathrm{MeV}$ with $10\%$ uncertainty, compared to the benchmarks $S_t$ (black line).  }
\label{fig:wave_m10_with_10}
\end{figure*}

We now present the reconstruction results  using the Tikhonov regularization given by Eq.\,(\ref{Tikhonov variation}). The recovered source functions for the potential $V_0=-10\,\mathrm{MeV}$ are shown in   Fig.\,\ref{fig:wave_m10_with_01} and Fig.\,\ref{fig:wave_m10_with_10}, corresponding to   $1\%$ and $10\%$,  respectively. In each subfigure, the black curve denotes the benchmark source $S_t$, the red curve represents the central reconstructed solution $S_\alpha^\epsilon$, and the pink shaded region indicates the associated error band. This band represents the uncertainty estimated by generating multiple datasets through random sampling within the experimental error margins, inverting each independently, and computing one standard deviation around the mean of the resulting ensemble of source functions. We observe that the reconstructed solutions for both single- and mixed-Gaussian source functions with $1\%$ uncertainty agree well with the benchmark.  Notably, the retrieved $S^\epsilon_\alpha $ naturally satisfies the normalization condition, with the integral approximately equal to unity. However, slight oscillations appear at large radii, as shown in Fig.~\ref{fig:wave_m10_with_01}.  These oscillations are attributed to the rapid decay of the wave function in that region, which reduces the distinguishability of the source function.  Although they lead to unphysical negative values in the reconstructed source, such artifacts are not a major concern here, as our primary focus is on the small-radii behavior, where the reconstruction is reliable. In future work, these issues could be further mitigated by incorporating additional prior information, such as non-negativity constraints.

\begin{figure*}[htbp]
\centering
\includegraphics[width=0.65\linewidth]{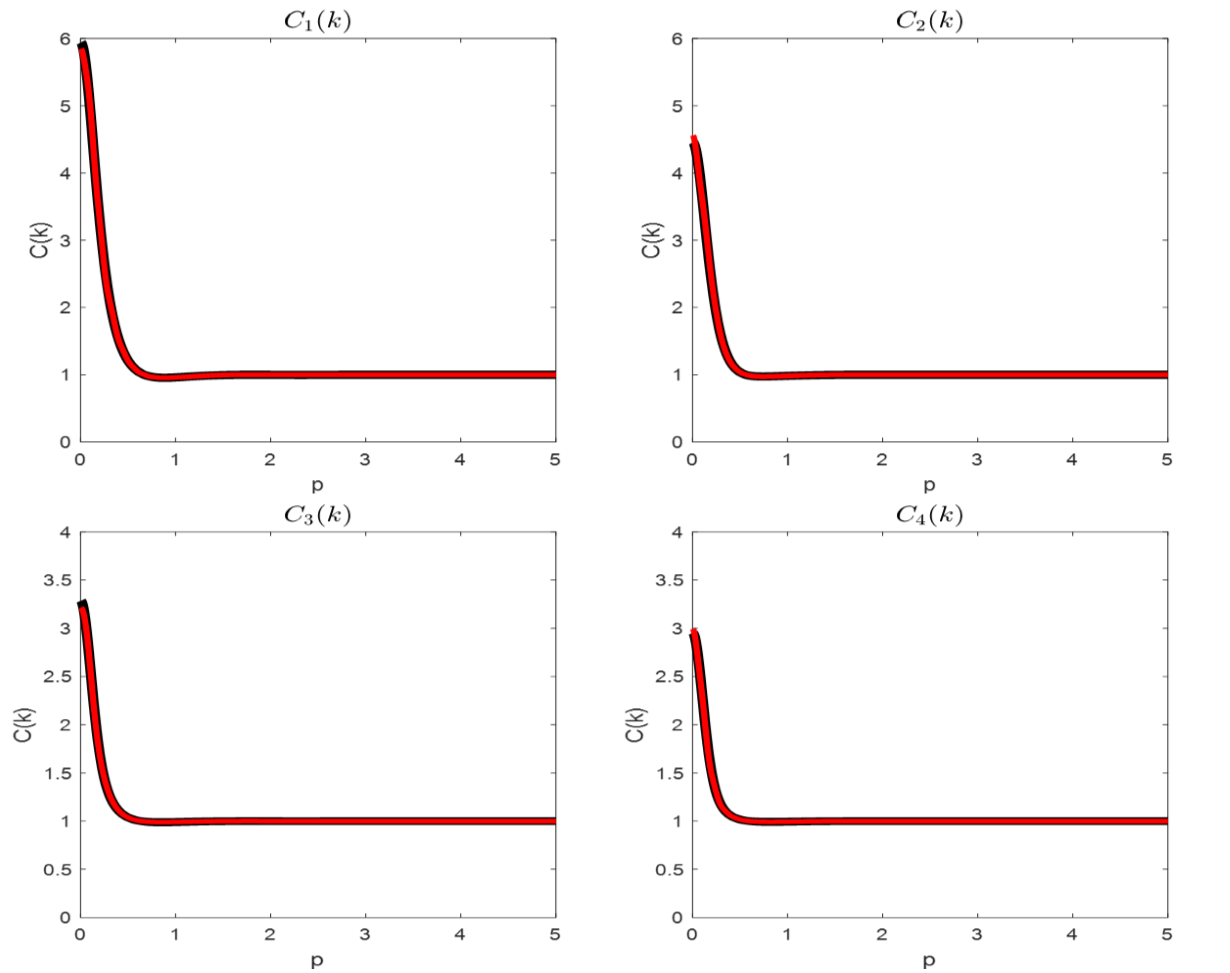}
\caption{Based on the KP formula, for $V_0=-10\,\mathrm{MeV}$ with $10\%$ uncertainty, the correlation function obtained from the central value of the reconstructed source $S_\alpha^\epsilon$ and the wave-function integral (red curve) is compared with that derived from the benchmark source $S_t$ (black curve). }
\label{fig:CFsm10_with_10}
\end{figure*}

In realistic experiments,  CF uncertainties often exceed  $1\%$. We therefore also consider the case with $10\%$ uncertainty.  As illustrated in Fig.~\ref{fig:wave_m10_with_10}, the reconstructed solutions for the single Gaussian source still match the benchmark well near the peak,  while some deviations occur for the mixed Gaussian source. This suggests that our method encounters challenges when applied to source functions with multiple peaks. Furthermore, the results indicate that reconstruction accuracy depends on the precision of the CFs.  This behavior aligns with the theoretical foundation of the Tikhonov regularization, as expressed in Eq.~(\ref{Tik_convergence}): as the input error decreases, the regularized solution converges to the true solution. To further validate our approach, we compute CFs using the reconstructed sources and the corresponding wave functions. For a representative case ($V_0=-10~\mathrm{MeV}$), we compare these with the true input CFs.  As shown in Fig.~\ref{fig:CFsm10_with_10}, the CFs obtained from the central values of the reconstructed sources, combined with the associated wave functions (black curves), closely match the true input CFs (red curves). This   agreement confirms the stability and reliability of our inverse approach.  
 
\begin{figure*}[htbp]
\centering
\includegraphics[width=0.9\linewidth]{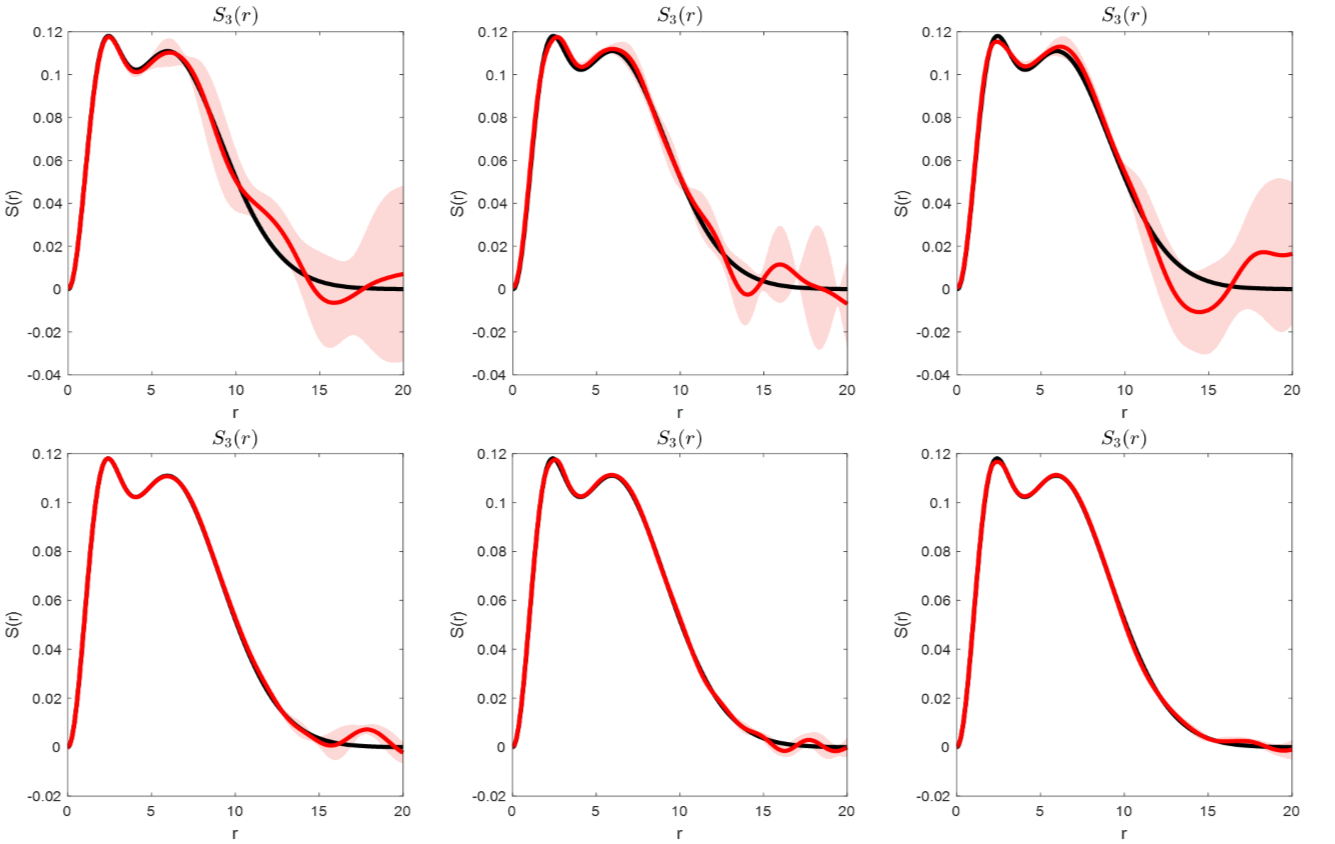}
\caption{Reconstruction of the source $S_3$ under three potential $25,-25$ and $-75\,\mathrm{MeV}$ with $10\%$ (top row) and $1\%$ (bottom row) uncertainty. Reconstructed solutions $S_\alpha^\epsilon$ (red lines) and the associated error bands (pink shaded areas), compared to the benchmarks $S_t$ (black line).}
\label{fig:different_potential_for_the_S3}
\end{figure*}

To investigate the general behavior under varying potential strengths, we examine the method by reconstructing the source $S_3(r,R)$ for three potentials: $ 25$, $-25$, and $-75$~MeV. As shown in Fig.~\ref{fig:wave_m10_with_10},  the behavior of the mixed source  $S_3$ is more complex than that of the other sources, making it a particularly stringent test for the Tikhonov regularization. In Fig.\,\ref{fig:different_potential_for_the_S3}, the reconstruction results under $10\%$ and $1\%$ uncertainties  are  presented  in the  top  and bottom panels, respectively.  In all scenarios, the reconstructed source (red curves) align closely with the benchmark solutions (black curves). These findings affirm the effectiveness and versatility of the Tikhonov regularization approach in reconstructing different source functions across a range of interaction potentials.  In short, the Tikhonov regularization method achieves consistent accuracy regardless of whether the true source is a single Gaussian or a mixture of Gaussians, demonstrating its ability to reconstruct source functions with complex structures reliably.

\section{Conclusions}

Momentum CFs are established as one effective physical observable for probing hadron-hadron interactions. Conventional approaches often assume a Gaussian parameterization of the source function to infer these interactions from CFs. However, the exact form of the source function is not known a priori. Reconstructing the source function from known momentum CFs is formally an inverse problem. In this study, we introduced a mathematically rigorous approach to solve this inverse problem.    In our calculations, we employed square-well potentials of four different strengths (repulsive, weakly attractive, moderately attractive, and deeply attractive) to derive analytic wave functions by solving the Schr\"odinger equation. Using these wave functions together with Gaussian and mixed Gaussian source functions, we generated the corresponding momentum CFs. We then reconstructed the source functions from the momentum CFs using the Tikhonov regularization.  

Our results show that the source functions for all four scenarios are well reproduced. Moreover, both the single- and mixed-Gaussian source functions can be well reproduced. We resampled the momentum CFs with introduced errors of $1\%$ and $10\%$. This revealed that the accuracy of the reproduced source functions depends on the precision of the momentum CFs; as the uncertainties in the input CFs increase, the fidelity of the reproduced momentum CFs degrades. Our results confirm that source functions can be reliably reconstructed via the Tikhonov regularization when both wave functions and momentum CFs are known. This approach paves the way for extracting realistic source functions for various hadron pairs—such as meson–meson, meson–baryon, and baryon–baryon systems—in the future, as precise hadron–hadron interactions and experimental momentum CFs become available.  Such source functions will be crucial for uncovering underlying physical properties and, in turn, will substantially improve the accuracy of extracted hadron–hadron interactions from momentum CFs.

\section{Acknowledgments}
We thank Ting Wei and Xiong-Bin Yan for valuable discussions in mathematics. This work is partly supported by the Fundamental and Interdisciplinary Disciplines Breakthrough Plan of the Ministry of Education of China-JYB2025XDXM204 and the National Natural Science Foundation of China under Grant No. W2543006 and No. 12435007. Ming-Zhu Liu acknowledges support from the National Natural Science Foundation of China under Grant No.12575086. Zhi-Wei Liu acknowledges support from the National Natural Science Foundation of China under Grant No.12405133, No.12347180, China Postdoctoral Science Foundation under Grant No.2023M740189, and the Postdoctoral Fellowship Program of CPSF under Grant No.GZC20233381. Fu-Sheng Yu and Ao-Sheng Xiong acknowledge support from the Scientific Research Innovation Capability Support Project for Young Faculty under Grant No. ZYGXQNJSKYCXNLZCXM-P2, and the Fundamental Research Funds for the Central Universities under No.~lzujbky-2023-stlt01, lzujbky-2024-oy02 and lzujbky-2025-eyt01.

\end{document}